\documentclass[11pt]{article}

\usepackage[T1]{fontenc}
\usepackage[utf8]{inputenc}
\usepackage[a4paper,margin=1in]{geometry}
\usepackage{amsmath,amssymb,bm,mathtools}
\usepackage{graphicx}
\usepackage{booktabs}
\usepackage{microtype}
\usepackage[numbers,sort&compress]{natbib}
\usepackage[hidelinks]{hyperref}
\usepackage[font=small,labelfont=bf]{caption}
\usepackage{xcolor}
\usepackage{enumitem}

\usepackage[normalem]{ulem}

\hypersetup{
    colorlinks=true,
    linkcolor=blue,
    citecolor=blue,
    urlcolor=blue
}

\makeatletter
\AtBeginDocument{%
  \let\Hy@oldcolorlink\Hy@colorlink
  \let\Hy@oldendcolorlink\Hy@endcolorlink
  \def\Hy@colorlink#1{\Hy@oldcolorlink{#1}\uline\bgroup}%
  \def\Hy@endcolorlink{\egroup\Hy@oldendcolorlink}%
}
\makeatother

\newcommand{\RR}{\mathbf{R}}
\newcommand{\rr}{\mathbf{r}}
\newcommand{\vv}{\mathbf{v}}
\newcommand{\FF}{\mathbf{F}}
\newcommand{\WW}{\mathbf{W}}
\newcommand{\Rgamma}{\mathbf{R}_{\gamma}}
\newcommand{\Ree}{R_{\mathrm{ee}}}
\newcommand{\Wi}{\mathrm{Wi}}
\newcommand{\Wref}{\mathrm{Wi}_{\mathrm{ref}}}
\newcommand{\Wrho}{\mathrm{Wi}_{\rho}}
\newcommand{\dd}{\mathrm{d}}

\title{\textbf{Blockwise friction heterogeneity redistributes stretching in a 2D Gaussian polymer under Batchelor-Kraichnan random flow}}
\author{Arpan Dey\thanks{Corresponding author. Email: \texttt{arpand2004@gmail.com}}\\
\small Universit\'e de Montpellier, Montpellier, France}
\date{}

\begin{document}
\maketitle

\begin{abstract}
We study a two-dimensional Gaussian Rouse chain divided into two contiguous blocks with different local friction coefficients but identical Hookean elasticity. In quiescent conditions, friction heterogeneity leaves the continuous-time equilibrium Gaussian conformation unchanged while producing distinct block mobilities and a modified generalized relaxation spectrum. The exact friction-weighted translational diffusion coefficient retains the scaling $D_\gamma\propto N^{-1}$, with a friction-dependent prefactor; the corresponding diffusion law and generalized-mode formulation extend beyond the diblock to more general local friction profiles. We then place the chain in an incompressible Batchelor-Kraichnan random flow. Under a common external flow, the high-friction block becomes more extended than the low-friction block. The deformation asymmetry strengthens with friction contrast and develops across the weak- to moderate-flow regime, remaining clearly visible at the strongest finite-time conditions. It also persists when the chain-specific longest-mode Weissenberg number is matched across friction ratios, showing that, within the same finite-time protocol, the effect is not accounted for solely by a shift of the global relaxation timescale. In generalized-mode coordinates, the smooth random strain does not directly mix mode indices and preferentially amplifies the slowest mode. At larger flow strengths, extension distributions broaden markedly and exhibit pronounced temporal drift over the simulated time window. Since all simulated chain-specific longest-mode Weissenberg numbers remain below unity, this drift is interpreted as slow convergence toward the broad, power-law-tailed stationary Hookean regime near the coil-stretch threshold. Overall, friction heterogeneity alone is sufficient to redistribute deformation along an otherwise mechanically uniform polymer.
\end{abstract}

\section{Introduction}

The Rouse model is the standard minimal description of the overdamped dynamics of a flexible polymer: a chain of beads connected by harmonic springs, with local viscous drag and thermal noise \cite{Rouse1953}. In its classical form, it deliberately neglects excluded-volume interactions, hydrodynamic interactions, bending stiffness, and finite-extensibility effects, thereby isolating the collective relaxation generated by chain connectivity. Real polymeric systems, however, need not be dynamically homogeneous along the backbone. Segmental mobility can vary through differences in local environment, compositional heterogeneity, or coupling of only part of a chain to additional nonequilibrium forcing. Generalizations of the Rouse model have therefore considered distributions
of segmental mobilities \cite{Hung2018}, random heterogeneous environments
\cite{Niedzwiedz2007}, active Rouse chains \cite{OsmanovicRabin2017}, and
spatially localized active forcing \cite{Osmanovic2018,Vatin2024}.

Polymers in random and turbulent flows provide a complementary setting in which the competition between polymer relaxation and fluctuating velocity gradients can be studied in a controlled way. Early studies examined Rouse chains in random convective environments \cite{OshaninBlumen1994}, while the Batchelor-Kraichnan framework and related smooth random-flow models have become standard reduced descriptions of small-scale stretching \cite{Batchelor1959,Kraichnan1968,Falkovich2001,Celani2005}. For elastic polymers, this competition gives rise to broad extension statistics and the coil-stretch transition \cite{Chertkov2000,Balkovsky2000,Celani2005,Gerashchenko2005,Celani2006}. More detailed bead-spring simulations in resolved turbulent flows have shown that chain length, coherent vortical structures, extreme strain events, finite extensibility, and hydrodynamic interactions can all modify polymer stretching \cite{Picardo2020,Picardo2023,Ganesh2026}. This problem is also closely connected to polymer-turbulence coupling and drag reduction \cite{WhiteMungal2008}.

The present work addresses the intersection of these two problems: how does a polymer with \emph{spatial friction heterogeneity} but uniform elasticity partition deformation under a common fluctuating flow? Here the different blocks of the polymer do not have different spring constants, contour lengths, or interaction energies. Instead, they experience different local friction coefficients, so the heterogeneity is kinetic and environmental rather than mechanical. Consequently, the equilibrium Boltzmann distribution is unchanged by the friction contrast, whereas the relaxation spectrum is not.

This construction is the continuum counterpart of our earlier dynamic Monte Carlo study of a two-dimensional Gaussian lattice polymer \cite{Dey2026}. In that model, the polymer was divided into two blocks that followed the same local bond-preserving move rules but were assigned different update rates. The higher-rate block displayed a larger early- and intermediate-time mean-squared displacement (MSD), while the full-chain center-of-mass diffusion retained the Rouse scaling $D_{\rm cm}\sim N^{-1}$. Here we make the corresponding friction field explicit in a continuous Langevin model and then use it to study flow-induced deformation. We first establish the quiescent benchmark, including the generalized relaxation spectrum and an exact friction-weighted translational diffusion law that extends directly to fixed multiblock friction profiles. We then add a two-dimensional incompressible Batchelor-Kraichnan flow and show that the high-friction block becomes preferentially more extended. The effect survives a control at matched chain-specific longest-mode Weissenberg number and therefore cannot be attributed solely to the increase of the slowest relaxation time.

\section{Heterogeneous Gaussian Rouse chain and quiescent benchmark}

\subsection{The model}

We consider an ideal Gaussian chain of $N$ beads in two dimensions. The first $N/2$ beads form block A and the remaining $N/2$ beads form block B. The harmonic energy is
\begin{equation}
U=\frac{k}{2}\sum_{i=1}^{N-1}\left|\rr_{i+1}-\rr_i\right|^2,
\label{eq:energy}
\end{equation}
with a \emph{single} spring constant $k$ throughout the chain. There is no excluded volume, bending stiffness, hydrodynamic interaction, or block-dependent elastic constant. The only heterogeneity is the bead friction
\begin{equation}
\gamma_i=
\begin{cases}
\gamma_A, & 1\le i\le N/2,\\
\gamma_B, & N/2<i\le N.
\end{cases}
\label{eq:gamma-profile}
\end{equation}

We characterize the dynamical heterogeneity by the friction ratio
\begin{equation}
\rho=\frac{\gamma_B}{\gamma_A},
\label{eq:rho-friction}
\end{equation}
with $\rho=1$ corresponding to a homogeneous chain. To vary the friction contrast while keeping the mean bead mobility fixed, we impose
\begin{equation}
\frac{1}{2}\left(\frac{1}{\gamma_A}+\frac{1}{\gamma_B}\right)
=\frac{1}{\gamma_0},
\label{eq:mean-mobility}
\end{equation}
where $\gamma_0$ is the friction coefficient of the corresponding homogeneous chain. Combining Eqs.~\eqref{eq:rho-friction} and \eqref{eq:mean-mobility} gives
\begin{equation}
\gamma_A=\gamma_0\frac{1+\rho}{2\rho},
\qquad
\gamma_B=\gamma_0\frac{1+\rho}{2}.
\label{eq:gammamap}
\end{equation}
Thus, for $\rho>1$, block A is the lower-friction, higher-mobility block, whereas block B is the higher-friction, lower-mobility block. The spring constant remains uniform throughout the chain, so the heterogeneity is purely dynamical: changing $\rho$ modifies the relaxation dynamics without altering the harmonic energy in Eq.~\eqref{eq:energy} or the equilibrium Boltzmann distribution.

For an interior bead the spring force is
\begin{equation}
\FF_i=k\left(\rr_{i+1}+\rr_{i-1}-2\rr_i\right),
\end{equation}
with the usual one-spring forces at the two free ends. In the absence of flow, the overdamped It\^o stochastic differential equation (SDE) is
\begin{equation}
\dd\rr_i=\frac{\FF_i}{\gamma_i}\,\dd t+
\sqrt{\frac{2k_{\rm B}T}{\gamma_i}}\,\dd\WW_i,
\label{eq:noflow-sde}
\end{equation}
where $\dd\WW_i$ are independent $d$-dimensional Wiener increments for
different beads. Equation~\eqref{eq:noflow-sde} satisfies the
fluctuation-dissipation relation locally. Since the friction coefficients do not enter the potential energy, the equilibrium distribution remains $P_{\rm eq}\propto \exp[-U/(k_{\rm B}T)]$ for every $\rho$.

\begin{figure}[t]
\centering
\includegraphics[width=0.48\linewidth]{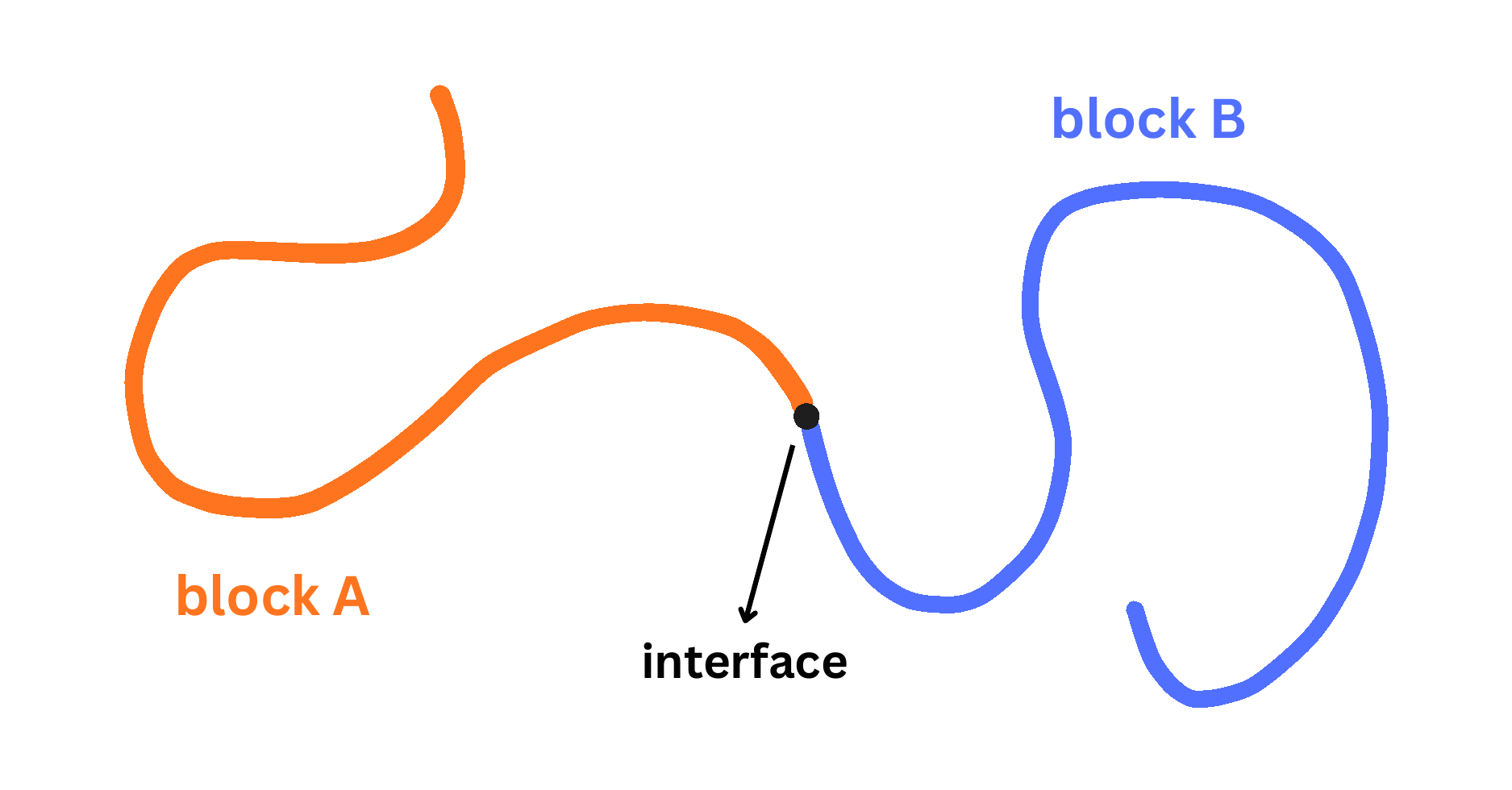}
\caption{Schematic of the heterogeneous Gaussian Rouse chain. The polymer consists of two equal contiguous blocks, A and B, connected at a single interface. The Hookean spring constant $k$ is uniform along the chain, while the friction coefficient is piecewise constant: $\gamma_A$ on block A and $\gamma_B$ on block B. Thus the heterogeneity is purely dynamical, entering only through the bead frictions and not through the elastic energy.}
\label{fig:model}
\end{figure}

\subsection{Relaxation modes and translational diffusion}

For one Cartesian coordinate, define the free-chain connectivity matrix $L$ and the diagonal friction matrix $\Gamma=\mathrm{diag}(\gamma_1,\ldots,\gamma_N)$. The deterministic dynamics is
\begin{equation}
\Gamma\dot{\mathbf{x}}=-kL\mathbf{x}.
\end{equation}

The relaxation modes are collective deformation patterns of the chain whose amplitudes decay exponentially in time. For mode $p$, $\lambda_p$ denotes the corresponding decay rate, so smaller $\lambda_p$ corresponds to slower relaxation.

The relaxation modes satisfy the generalized eigenvalue problem
\begin{equation}
kL\vv_p=\lambda_p\Gamma\vv_p,
\qquad
\vv_p^{\mathsf T}\Gamma\vv_q=\delta_{pq}.
\label{eq:generalized-eigen}
\end{equation}
Here $\vv_p$ specifies the bead-wise shape of mode $p$, while the second relation gives the friction-weighted orthonormality of the generalized modes. The generalized problem can be converted to the ordinary eigenvalue problem for the matrix
\begin{equation*}
k\Gamma^{-1/2}L\Gamma^{-1/2},
\end{equation*}
which is real symmetric because both $L$ and the diagonal matrix $\Gamma^{-1/2}$ are symmetric. Its eigenvalues are therefore real; moreover, since $L$ is positive semidefinite, all $\lambda_p$ are non-negative. The zero eigenvalue corresponds to rigid translation of the entire chain, which leaves all bond extensions unchanged. The smallest nonzero eigenvalue, $\lambda_1$, sets the slowest internal relaxation time,
\begin{equation}
\tau_1=\lambda_1^{-1}.
\end{equation}

Within each block, where $\gamma_i$ is constant, the mode equation reduces to a second-order difference equation, the discrete analogue of the Helmholtz equation. For a homogeneous segment with friction $\gamma_\alpha$, oscillatory mode shapes are possible only for eigenvalues in the range
\begin{equation}
0\leq \lambda \leq \frac{4k}{\gamma_\alpha}.
\end{equation}
This interval defines the local spectral band of that block. Because the two blocks have different frictions, their spectral-band limits are different. A global eigenmode may therefore remain oscillatory in one block while becoming spatially decaying in the other when its eigenvalue exceeds the latter block's band edge. The friction discontinuity can thus reshape the global eigenmodes and modify the relaxation spectrum even though the spring matrix $L$ itself is unchanged.

A particularly useful coordinate is the friction-weighted center
\begin{equation}
\Rgamma=\frac{\sum_i\gamma_i\rr_i}{\sum_i\gamma_i}.
\label{eq:Rgamma}
\end{equation}

For comparison, the ordinary arithmetic center of mass is
\begin{equation}
\RR_{\rm cm}
=
\frac{1}{N}\sum_{i=1}^{N}\rr_i.
\end{equation}

Multiplying Eq.~\eqref{eq:noflow-sde} for each bead by its friction coefficient $\gamma_i$ and summing over the chain causes the internal spring forces to cancel pairwise. The resulting equation describes the Brownian motion of the friction-weighted center and gives the exact diffusion coefficient
\begin{equation}
D_\gamma=\frac{k_{\rm B}T}{\sum_i\gamma_i}
=\frac{k_{\rm B}T}{\gamma_0}\frac{4\rho}{N(1+\rho)^2},
\label{eq:Dgamma}
\end{equation}
for two equal blocks. Thus the friction contrast changes the diffusion prefactor through $\rho$ while preserving the analytic chain-length dependence $D_\gamma\propto N^{-1}$ at fixed block fractions. In the present $N=100$ no-flow Langevin simulations, the measured translational diffusion coefficients follow the expected $\rho$-dependence and remain close to the exact values predicted by Eq.~\eqref{eq:Dgamma} for all three friction contrasts. At finite times, the ordinary arithmetic center of mass can contain contributions from internal relaxation modes, whereas $\Rgamma$ isolates the translational zero mode exactly. After the internal modes have relaxed, the ordinary center of mass acquires the same long-time diffusive slope and therefore the same diffusion coefficient $D_\gamma$.

More generally, the translational result is not restricted to the equal
diblock geometry. Consider a chain divided into $M$ blocks, with block
$\alpha$ containing a fixed fraction $f_\alpha=N_\alpha/N$ of the beads
and having friction coefficient $\gamma_\alpha$, where
$\sum_{\alpha=1}^{M}f_\alpha=1$. The total friction is then
\begin{equation}
\sum_{i=1}^{N}\gamma_i
=
N\sum_{\alpha=1}^{M}f_\alpha\gamma_\alpha,
\end{equation}
and the exact friction-weighted diffusion coefficient becomes
\begin{equation}
D_\gamma
=
\frac{k_{\rm B}T}
{N\sum_{\alpha=1}^{M}f_\alpha\gamma_\alpha}.
\end{equation}
Thus any fixed multiblock friction profile preserves the chain-length
scaling $D_\gamma\propto N^{-1}$, while changing its prefactor. If the
mean bead mobility is held fixed according to
$\sum_{\alpha}f_\alpha/\gamma_\alpha=1/\gamma_0$, the weighted
arithmetic-harmonic mean inequality further gives
\begin{equation}
D_\gamma
\leq
\frac{k_{\rm B}T}{N\gamma_0},
\end{equation}
with equality only for the homogeneous chain. Hence, under this
normalization, friction heterogeneity can only reduce the translational
diffusion prefactor relative to the homogeneous reference while leaving
the $N^{-1}$ scaling unchanged.

\subsection{Numerical quiescent benchmark}

The no-flow SDE is integrated with fixed-step Euler-Maruyama. Because the drift is linear and the thermal noise is additive, the generalized internal modes are Ornstein-Uhlenbeck processes; however, a finite Euler-Maruyama timestep does not reproduce their continuous-time stationary covariance exactly. We therefore retain the simple fixed-step scheme used here but quantify its known $\mathcal O(\Delta t)$ equilibrium-covariance bias below. We use dimensionless units $k_{\rm B}T=1$, $\gamma_0=1$, and choose the continuous-time equilibrium mean-squared bond length $\left\langle |\mathbf b|^2\right\rangle$ to be unity. For the Gaussian spring potential in dimension $d$ this gives
\[
\left\langle |\mathbf b|^2\right\rangle
=\frac{d k_{\rm B}T}{k},
\]
so that in two dimensions the corresponding spring constant is $k=2$. The principal benchmark uses $N=100$ beads in $d=2$, $\rho=1,2,4$, $100$ independent realizations, a time step $\Delta t=0.01$, and a total duration of $t_{\rm tot}=10^4$. Each realization is initialized as an independent equilibrium Gaussian chain generated directly from the exact continuous-time bond distribution. This avoids a separate preliminary equilibration stage, although the subsequent finite-step Euler-Maruyama dynamics relaxes toward the nearby discrete-time stationary covariance discussed below.

\begin{figure}[t]
\centering
\includegraphics[width=\textwidth]{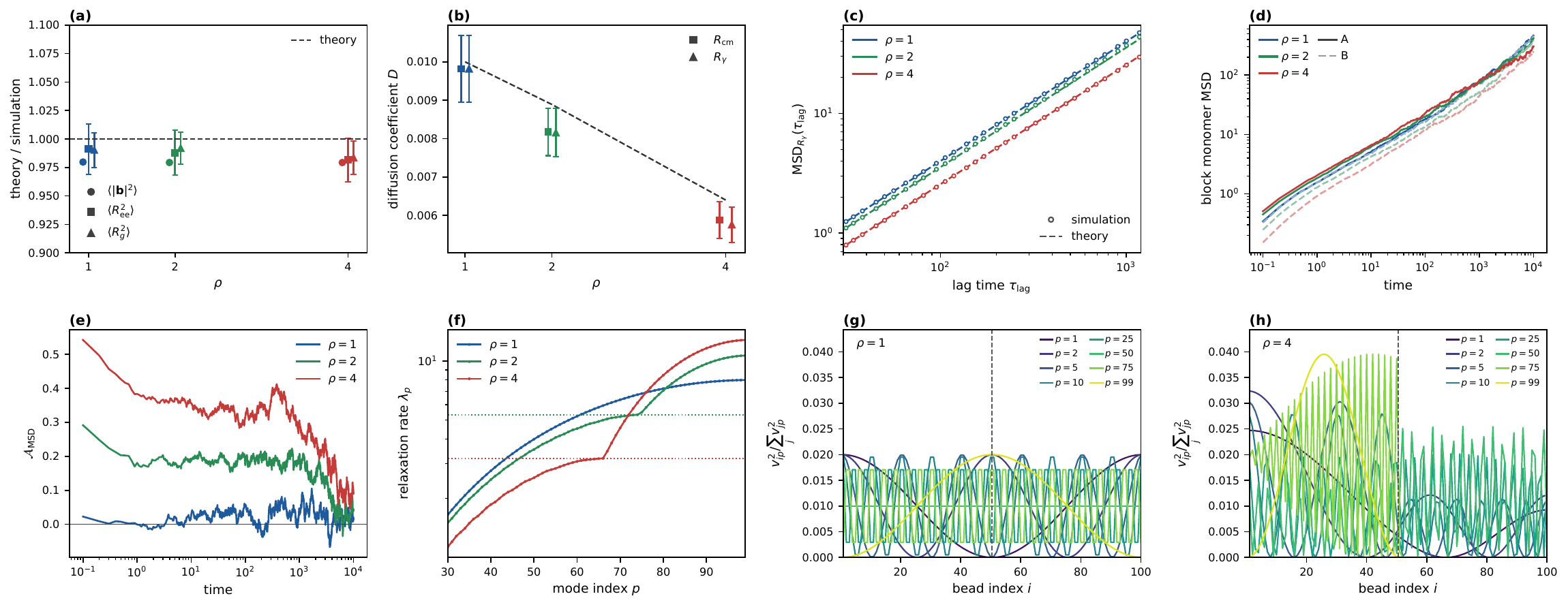}
\caption{
Quiescent benchmark of the continuous heterogeneous Rouse model for $N=100$ and $\rho=1,2,4$.
(a) Ratios of continuous-time theoretical to simulated values of the mean-squared bond length $\left\langle |\mathbf b|^2\right\rangle$, end-to-end distance $\langle R_{\rm ee}^2\rangle$, and radius of gyration $\langle R_g^2\rangle$; the dashed line denotes perfect agreement. The approximately $2\%$ offset of the bond statistic is the expected finite-$\Delta t$ Euler-Maruyama covariance bias.
(b) Long-time diffusion coefficients obtained from the ordinary center of mass $\RR_{\rm cm}$ and the friction-weighted center $\Rgamma$, compared with the theoretical prediction $D_\gamma=k_{\rm B}T/\sum_i\gamma_i$; the dashed segments connect the three theoretical values as a guide to the eye.
(c) Time-averaged MSD of $\Rgamma$ versus lag time $\tau_{\rm lag}$; circles denote simulation data and dashed lines the Brownian prediction ${\rm MSD}_{R_\gamma}=2dD_\gamma\tau_{\rm lag}$ in $d=2$.
(d) Block-resolved monomer MSDs, with solid and dashed curves denoting blocks A and B, respectively.
(e) Block mobility asymmetry $\mathcal{A}_{\rm MSD}=(\mathrm{MSD}_A-\mathrm{MSD}_B)/(\mathrm{MSD}_A+\mathrm{MSD}_B)$.
(f) High-mode sector of the generalized relaxation spectrum $\lambda_p$; dotted horizontal lines mark the local spectral-band edge $4k/\gamma_B$ of the high-friction block B.
(g) and (h) Selected normalized eigenvector weights $v_{ip}^2/\sum_jv_{jp}^2$ for $\rho=1$ and $\rho=4$, respectively; the vertical dashed line marks the block interface.
Blue, green, and red denote $\rho=1,2,4$, respectively. Error bars in panels (a) and (b) are SEMs across independent realizations.
}
\label{fig:noflow}
\end{figure}

The equilibrium predictions are
\begin{equation}
\left\langle |\mathbf b|^2\right\rangle=1,
\qquad
\left\langle R_{\rm ee}^2\right\rangle=N-1,
\qquad
\left\langle R_g^2\right\rangle=\frac{N^2-1}{6N},
\label{eq:eqstats}
\end{equation}
where $\mathbf b_i=\mathbf r_{i+1}-\mathbf r_i$ is a bond vector,
$R_{\rm ee}=|\mathbf r_N-\mathbf r_1|$ is the end-to-end distance, and
$R_g$ is the radius of gyration. The latter may be written as
\begin{equation}
R_g^2
=
\frac{1}{2N^2}
\sum_{i=1}^{N}\sum_{j=1}^{N}
|\mathbf r_i-\mathbf r_j|^2,
\end{equation}
which is the form used in the analysis. For an ideal Gaussian chain,
$\langle |\mathbf r_i-\mathbf r_j|^2\rangle=|i-j|$ in the present units, yielding the finite-$N$ result in Eq.~\eqref{eq:eqstats}.

The continuous-time equilibrium quantities in Eq.~\eqref{eq:eqstats} are independent of $\rho$, since changing the friction coefficients modifies the dynamics but not the Boltzmann distribution. The small systematic offset visible in Fig.~\ref{fig:noflow}(a) for the bond statistic is a finite-timestep effect of the Euler-Maruyama discretization. For a generalized internal mode $p$, one Cartesian component obeys the discrete update
\begin{equation}
a_{p}^{n+1}
=
(1-\lambda_p\Delta t)a_{p}^{n}
+
\sqrt{2k_{\rm B}T\Delta t}\,\xi_p^n,
\end{equation}
so its discrete stationary variance is
\begin{equation}
\left\langle a_p^2\right\rangle_{\rm EM}
=
\frac{k_{\rm B}T/\lambda_p}
     {1-\lambda_p\Delta t/2}.
\label{eq:em-mode-bias}
\end{equation}
Thus rapidly relaxing modes are slightly overrepresented relative to the continuous-time equilibrium measure. At $\Delta t=0.01$, the resulting predictions are
$\langle |\mathbf b|^2\rangle_{\rm EM}\simeq1.0206$, $1.0207$, and $1.0209$ for $\rho=1,2,4$, respectively, which accounts quantitatively for the approximately $2\%$ theory-to-simulation offset in Fig.~\ref{fig:noflow}(a). By contrast, large-scale conformational observables are only weakly affected: for example, $\langle R_{\rm ee}^2\rangle_{\rm EM}\simeq99.02$ compared with the continuous-time value $99$.

The translational dynamics provides an independent dynamical benchmark. Figure~\ref{fig:noflow}(b) compares diffusion coefficients extracted from the long-time motion of $\RR_{\rm cm}$ and $\Rgamma$ with Eq.~\eqref{eq:Dgamma}; for $N=100$, the theoretical values are $D_\gamma=0.0100$, $0.00889$, and $0.00640$ for $\rho=1,2,4$, respectively. The fitted values in panel (b) lie slightly below the exact prediction in a common direction, with deviations comparable to the realization-to-realization uncertainty. This offset is not the Euler-Maruyama covariance bias discussed above: for $\Rgamma$ the internal forces cancel exactly at every discrete timestep, leaving a Gaussian Brownian increment with the exact diffusion coefficient $D_\gamma$. The small panel-(b) discrepancy is therefore consistent with finite-trajectory and fitting-window uncertainty. Figure~\ref{fig:noflow}(c) provides a more direct test through the time-averaged MSD of $\Rgamma$. Here $\tau_{\rm lag}$ denotes the temporal separation between two configurations along a trajectory, and the MSD is averaged over all available time origins at that separation. The numerical curves closely follow
\begin{equation}
{\rm MSD}_{R_\gamma}(\tau_{\rm lag})
=
2dD_\gamma\tau_{\rm lag},
\end{equation}
confirming both the diffusive scaling and the predicted $\rho$-dependent diffusion coefficient.

The internal bead dynamics reflects the imposed friction contrast directly. Figure~\ref{fig:noflow}(d) shows the block-resolved monomer MSDs. At intermediate times the connected chain exhibits the familiar subdiffusive Rouse regime, before crossing over at long times to collective translational diffusion. For $\rho>1$, the lower-friction block A has the larger MSD, whereas the higher-friction block B moves more slowly. The dependence on $\rho$ is especially pronounced in block B because $\gamma_B$ increases strongly with the friction contrast, reducing its local mobility. The corresponding separation among the block-A curves is weaker: although $\gamma_A$ decreases and block A becomes locally more mobile, its motion remains coupled through the chain connectivity to the increasingly slow block B. At sufficiently long times the distinction between the two blocks disappears as the common translational motion of the entire chain becomes dominant, with the overall diffusion coefficient controlled by Eq.~\eqref{eq:Dgamma}.

To quantify the relative block mobility, we define
\begin{equation}
\mathcal{A}_{\rm MSD}(t)
=
\frac{\mathrm{MSD}_A(t)-\mathrm{MSD}_B(t)}
{\mathrm{MSD}_A(t)+\mathrm{MSD}_B(t)}.
\label{eq:msd-asymmetry}
\end{equation}
As shown in Fig.~\ref{fig:noflow}(e), $\mathcal{A}_{\rm MSD}$ remains close to zero for the homogeneous chain, while it is positive for $\rho>1$ and increases with the friction contrast at early and intermediate times. Its subsequent decrease reflects the crossover from block-dependent internal motion to the common long-time translation of the whole chain. This behavior is consistent with the mobility contrast previously observed in the lattice model \cite{Dey2026}.

The continuum formulation additionally provides direct access to the heterogeneous relaxation spectrum. For $N=100$, direct diagonalization of the generalized Rouse problem gives the slowest internal relaxation times
\begin{equation}
\tau_1(\rho=1)=506.65,
\qquad
\tau_1(\rho=2)=523.12,
\qquad
\tau_1(\rho=4)=590.51.
\label{eq:tau-values}
\end{equation}
Thus increasing the friction contrast slows the longest collective relaxation, even though block A itself becomes progressively more mobile. The slowest modes extend over the connected chain and therefore remain strongly influenced by the increasingly high-friction block B.

The higher modes show a different reorganization, as shown in Fig.~\ref{fig:noflow}(f). The plotted eigenvalues $\lambda_p$ are relaxation rates of global modes of the entire connected chain. At low and intermediate mode index, these modes extend across both blocks, so increasing the friction of block B tends to reduce their relaxation rates. At larger $p$, however, the two local spectral-band limits become important. Within block $\alpha$, oscillatory mode structure is supported only up to
\[
\lambda_{\max,\alpha}=\frac{4k}{\gamma_\alpha}.
\]
For $\rho>1$, $\gamma_A<\gamma_B$, and therefore $\lambda_{\max,A}>\lambda_{\max,B}$. Once the eigenvalue of a global mode exceeds the B-block limit $4k/\gamma_B$, that mode can no longer remain oscillatory throughout block B and its amplitude becomes strongly attenuated there. The same eigenvalue can still be supported in the lower-friction block A, so the mode becomes increasingly concentrated in A. As this spatial redistribution occurs, the high-mode relaxation rates rise toward the larger A-block spectral range, producing the reversal of the spectral ordering visible in Fig.~\ref{fig:noflow}(f).

This redistribution is visible directly in Figs.~\ref{fig:noflow}(g) and \ref{fig:noflow}(h), where we plot the normalized eigenvector weight
\begin{equation}
\frac{v_{ip}^2}{\sum_j v_{jp}^2}   
\end{equation}

as a function of bead index $i$ for selected modes $p$. This quantity measures how the squared amplitude of a given eigenmode is distributed along the polymer backbone, with the normalization ensuring that the total plotted weight of each mode is unity. For the homogeneous chain, the selected generalized eigenmodes remain extended across both halves of the polymer, with no preferred block. At $\rho=4$, by contrast, the higher-frequency modes acquire strongly unequal weight across the interface and become concentrated predominantly in the low-friction block A, while their weight is suppressed in block B. The low-order modes remain comparatively extended because their eigenvalues lie well below the block spectral edges. The eigenvector-weight profiles therefore provide the spatial counterpart of the spectral reorganization observed in Fig.~\ref{fig:noflow}(f).

\section{Batchelor-Kraichnan flow}

\subsection{Spatially smooth random-flow model}

We next place the chain in a synthetic random flow in the spatially smooth Batchelor regime. Let $\mathbf u(\mathbf r,t)$ denote the solvent velocity field. When the polymer is small compared with the spatial scale over which $\mathbf u$ varies appreciably, the velocity field can be expanded to first order about the friction-weighted center $\Rgamma$. Retaining only the linear term gives
\begin{equation}
\mathbf u(\rr_i,t)-\mathbf u(\Rgamma,t)
=
\bm{\sigma}(t)\left(\rr_i-\Rgamma\right),
\label{eq:batchelor-linear}
\end{equation}
where
\begin{equation}
\sigma_{ij}(t)
=
\left.
\frac{\partial u_i}{\partial r_j}
\right|_{\mathbf r=\Rgamma}
\end{equation}
is the local velocity-gradient tensor. Equation~\eqref{eq:batchelor-linear} is therefore the linear, first-order Taylor approximation to the velocity difference across the polymer. The uniform velocity $\mathbf u(\Rgamma,t)$ advects every bead equally and hence changes no pair separation, end-to-end vector, radius of gyration, or internal mode. Working in the co-moving frame of $\Rgamma$ thus removes this irrelevant absolute translation.

We adopt the standard smooth incompressible Kraichnan ensemble for the velocity gradient \cite{Kraichnan1968,Falkovich2001,Celani2005}. The random tensor $\bm{\sigma}(t)$ has zero mean, Gaussian statistics, no preferred spatial direction, and no temporal correlation. Its covariance is
\begin{equation}
\left\langle\sigma_{ij}(t)\sigma_{kl}(t')\right\rangle
=
2D_1\left[
(d+1)\delta_{ik}\delta_{jl}
-\delta_{ij}\delta_{kl}
-\delta_{il}\delta_{jk}
\right]\delta(t-t').
\label{eq:kraichnan-cov}
\end{equation}
Here $D_1$ sets the strength of the velocity-gradient fluctuations, while the Dirac delta expresses the white-in-time assumption. The tensorial combination of Kronecker deltas is the standard isotropic incompressible form: it introduces no preferred direction and enforces
$\mathrm{Tr}\,\bm{\sigma}=0$, corresponding to $\nabla\cdot\mathbf u=0$.

The physical stretching rate associated with this random flow is characterized by its largest Lyapunov exponent. For an infinitesimal passive separation vector $\mathbf R(t)$, the linearized flow gives
\begin{equation}
\dot{\mathbf R}=\bm{\sigma}(t)\mathbf R.
\end{equation}
In a smooth chaotic flow with a positive largest Lyapunov exponent, infinitesimal separations grow asymptotically exponentially,
$|\mathbf R(t)|\sim |\mathbf R(0)|e^{\lambda_{\rm flow}t}$, with
\begin{equation}
\lambda_{\rm flow}
=
\lim_{t\to\infty}
\frac{1}{t}
\left\langle
\ln\frac{|\mathbf R(t)|}{|\mathbf R(0)|}
\right\rangle .
\label{eq:lyapunov-definition}
\end{equation}
For the isotropic incompressible Kraichnan ensemble defined by
Eq.~\eqref{eq:kraichnan-cov}, the separation dynamics can be solved exactly and gives
\begin{equation}
\lambda_{\rm flow}=D_1d(d-1).
\label{eq:flow-lyapunov}
\end{equation}
Thus, in the present two-dimensional system,
$\lambda_{\rm flow}=2D_1$. This quantity provides the characteristic stretching rate of the imposed random flow and will be used below to define the Weissenberg number.

To couple the heterogeneous Rouse chain to the random flow, we retain the usual spring relaxation and thermal Brownian motion while adding the relative solvent velocity across the chain. In the Batchelor approximation of Eq.~\eqref{eq:batchelor-linear}, the flow contribution is linear in the displacement from the friction-weighted center. In the white-in-time Kraichnan limit, the infinitesimal velocity-gradient contribution $\bm{\sigma}(t)\,\dd t$ is represented by a matrix Wiener increment $\dd\mathbf B(t)$, whose covariance is determined by Eq.~\eqref{eq:kraichnan-cov}. The resulting polymer dynamics is
\begin{equation}
\dd\rr_i=
\frac{\FF_i}{\gamma_i}\,\dd t
+\circ\dd\mathbf B(t)\left(\rr_i-\Rgamma\right)
+\sqrt{\frac{2k_{\rm B}T}{\gamma_i}}\,\dd\WW_i.
\label{eq:flow-sde}
\end{equation}
The three terms describe, respectively, elastic relaxation, deformation by the fluctuating velocity gradient, and thermal Brownian motion. The matrix increment satisfies the same isotropic and incompressible tensor structure as Eq.~\eqref{eq:kraichnan-cov}, with covariance proportional to $\dd t$. The circle denotes the Stratonovich interpretation of the multiplicative flow noise, corresponding to the short-correlation-time limit of a rapidly fluctuating smooth velocity gradient. The thermal noise is additive and independent of the imposed flow noise.

The competition between flow-induced stretching and polymer relaxation is characterized by a Weissenberg number. We distinguish a reference value,
\begin{equation}
\Wref=\lambda_{\rm flow}\tau_1(\rho=1),
\qquad
\Wrho=\lambda_{\rm flow}\tau_1(\rho),
\label{eq:Wi-def}
\end{equation}
where $\Wref$ measures the imposed flow strength relative to the relaxation time of the homogeneous reference chain, while $\Wrho$ is the corresponding value for a chain with friction contrast $\rho$. At fixed $\Wref$, $\lambda_{\rm flow}$, and hence $D_1$, is identical for all $\rho$, so the polymer is always subjected to the same random-flow statistics. Because the relaxation times differ,
\begin{equation}
\Wrho
=
\Wref\,
\frac{\tau_1(\rho)}{\tau_1(\rho=1)},
\end{equation}
and the heterogeneous chain therefore experiences different ratios of flow stretching to internal relaxation. In the primary comparison, we keep the external flow statistics fixed across the separate simulations at different $\rho$ and do not adjust $D_1$ to compensate for the $\rho$-dependence of $\tau_1$. A separate matched-$\Wrho$ control is then introduced, in which the flow strength is adjusted so that $\Wrho$ is the same for each friction contrast. This distinguishes effects caused by the change in the overall slowest relaxation time from those arising solely from the spatial distribution of friction along the chain.

\subsection{Modal dynamics in the smooth-flow limit}

The linear structure of the Batchelor-Kraichnan model allows the bead dynamics to be expressed directly in terms of the generalized Rouse modes introduced previously. Let $\vv_p$ denote the generalized no-flow eigenvector satisfying Eq.~\eqref{eq:generalized-eigen}. For each internal mode $p\geq 1$, we define its vector amplitude by
\begin{equation}
\mathbf a_p
=
\sum_{i=1}^{N}
\gamma_i v_{ip}
\left(\rr_i-\Rgamma\right).
\label{eq:modeamp}
\end{equation}
The coefficients $\gamma_i v_{ip}$ arise from the friction-weighted orthonormality of the generalized eigenvectors,
$\vv_p^{\mathsf T}\Gamma\vv_q=\delta_{pq}$. Thus Eq.~\eqref{eq:modeamp} is the natural projection of the instantaneous polymer conformation onto mode $p$. The subtraction of $\Rgamma$ removes rigid translation; equivalently, every internal mode is friction-weighted orthogonal to the translational zero mode and therefore satisfies
\begin{equation}
\sum_{i=1}^{N}\gamma_i v_{ip}=0,
\qquad p\geq 1.
\end{equation}
The vector $\mathbf a_p$ describes the instantaneous amplitude and orientation of mode $p$ in physical space, whereas $\vv_p$ specifies its bead-wise shape along the polymer backbone.

Projecting the bead-level stochastic equation, Eq.~\eqref{eq:flow-sde}, onto this generalized basis gives a particularly simple modal dynamics. The elastic contribution reduces to
$-\lambda_p\mathbf a_p\,\dd t$ by virtue of the generalized eigenvalue equation. Because the same velocity-gradient increment acts on every bead, $\dd\mathbf B(t)$ is independent of the bead index and can be factored out of the projection sum. The remaining sum is precisely the modal amplitude $\mathbf a_p$, so the flow contribution becomes $\circ\dd\mathbf B(t)\mathbf a_p$ and does not directly couple mode $p$ to other mode indices. The projected equation is
\begin{equation}
\dd\mathbf a_p
=
-\lambda_p\mathbf a_p\,\dd t
+\circ\dd\mathbf B(t)\,\mathbf a_p
+\sqrt{2k_{\rm B}T}\,\dd\mathbf W_p.
\label{eq:mode-sde}
\end{equation}
Here the projected thermal increment is
\begin{equation}
\dd\mathbf W_p
=
\sum_{i=1}^{N}
\sqrt{\gamma_i}\,v_{ip}\,
\dd\mathbf W_i.
\end{equation}
Because the bead-level thermal noises are independent and the generalized modes satisfy
$\vv_p^{\mathsf T}\Gamma\vv_q=\delta_{pq}$, these projected increments obey
\begin{equation}
\left\langle
\dd W_{p,\alpha}\,
\dd W_{q,\beta}
\right\rangle
=
\delta_{pq}\delta_{\alpha\beta}\,\dd t.
\end{equation}
Thus the thermal forcing remains diagonal in mode index.

Equation~\eqref{eq:mode-sde} is exact within the spatially linear Batchelor approximation used here. All internal modes experience the same realization of the random velocity gradient, but each relaxes with its own rate $\lambda_p$. Importantly, the flow term contains $\mathbf a_p$ itself and no amplitude $\mathbf a_q$ with $q\neq p$. The spatially uniform velocity gradient therefore produces no direct linear mixing between different generalized mode indices. The modes are nevertheless not statistically independent, because the same realization of $\dd\mathbf B(t)$ acts on all of them simultaneously.

The competition between random stretching and elastic relaxation can then be considered separately for each mode. Defining $\tau_p=\lambda_p^{-1}$, the corresponding mode-specific Weissenberg number is
\begin{equation}
\Wi_p
=
\lambda_{\rm flow}\tau_p
=
\frac{\lambda_{\rm flow}}{\lambda_p}.
\label{eq:Wi-mode}
\end{equation}
For a fixed external flow, slower modes have larger $\tau_p$ and therefore larger $\Wi_p$. They are consequently more susceptible to amplification by the random velocity gradient, whereas rapidly relaxing higher modes have smaller $\Wi_p$.

To see the competition directly, consider Eq.~\eqref{eq:mode-sde} without the additive thermal-noise term. The random flow tends to stretch mode $p$ at the rate $\lambda_{\rm flow}$, while elastic relaxation tends to reduce its amplitude at the rate $\lambda_p$. The resulting net exponential growth rate is therefore
\begin{equation}
\lambda_{\rm flow}-\lambda_p.
\end{equation}
The two effects balance when
$\lambda_{\rm flow}=\lambda_p$,
or equivalently when $\Wi_p=1$. This provides the mode-level analogue of the competition between random-flow stretching and elastic relaxation familiar from coil-stretch theory
\cite{Chertkov2000,Balkovsky2000,Celani2005}.

For the Kraichnan ensemble, the same modal equation also determines the nature of the stationary extension statistics below this threshold. Let $A_p=|\mathbf a_p|$. For $\Wi_p<1$, the radial stationary density of a Hookean mode is normalizable and has the asymptotic large-amplitude form
\begin{equation}
p_p(A_p)
\sim
A_p^{-1-q_p},
\qquad
q_p
=
d\left(\frac{1}{\Wi_p}-1\right),
\label{eq:mode-tail}
\end{equation}
whereas no normalizable Hookean stationary density exists for $\Wi_p\geq1$ \cite{Celani2005}. The $m$th moment exists only for $m<q_p$. In the present two-dimensional system this implies that the stationary second moment already diverges for $\Wi_p\geq1/2$ and the stationary first moment diverges for $\Wi_p\geq2/3$, even though the probability density itself remains normalizable until $\Wi_p=1$. This distinction is important below: close to the coil-stretch threshold, medians remain well-defined and numerically much more robust than moments that are dominated by the algebraic tail. 

The generalized-mode construction is also not restricted to the
equal diblock geometry used in the simulations. For any positive
diagonal friction profile
$\Gamma=\mathrm{diag}(\gamma_1,\ldots,\gamma_N)$, the generalized
eigenproblem
$kL\mathbf v_p=\lambda_p\Gamma\mathbf v_p$ and the projection leading
to Eq.~\eqref{eq:mode-sde} retain the same form. Consequently, a
spatially uniform Batchelor-Kraichnan velocity gradient remains
diagonal in the generalized mode index: the friction profile changes
the relaxation rates and mode shapes, but does not introduce direct
linear coupling between different generalized modes. For a
piecewise-constant multiblock chain, each block $\alpha$ additionally
has the local spectral edge
\begin{equation*}
\lambda_{\max,\alpha}=\frac{4k}{\gamma_\alpha},
\end{equation*}
so a sequence of distinct block frictions produces a corresponding
hierarchy of local spectral cutoffs. Modes of sufficiently large
$\lambda_p$ can therefore remain oscillatory in lower-friction blocks
while becoming spatially attenuated in higher-friction blocks. 

\subsection{Numerical protocol and diagnostics}

The multiplicative white-in-time flow term is integrated in the
Stratonovich sense using a stochastic Heun predictor-corrector scheme.
Let $\rr_i^n$ denote the position of bead $i$ at timestep $n$. A
provisional configuration is first constructed as
\begin{equation}
\rr_i^*
=
\rr_i^n
+\frac{\FF_i^n}{\gamma_i}\Delta t
+\Delta\mathbf B^n
\left(\rr_i^n-\Rgamma^n\right)
+\Delta\rr_{i,\rm th},
\label{eq:heun-predictor}
\end{equation}
where $\FF_i^n$ and $\Rgamma^n$ are evaluated from the current
configuration. The thermal displacement is
\begin{equation}
\Delta\rr_{i,\rm th}
=
\sqrt{\frac{2k_{\rm B}T\Delta t}{\gamma_i}}\,
\bm{\xi}_i^n,
\qquad
\bm{\xi}_i^n\sim\mathcal N(\mathbf 0,\mathbf I_d),
\label{eq:thermal-increment}
\end{equation}
with independent Gaussian vectors for different beads and timesteps.
The thermal increments are also statistically independent of the
random-flow increments.

The matrix $\Delta\mathbf B^n$ represents the velocity-gradient
increment accumulated during the same timestep and satisfies
\begin{equation}
\left\langle
\Delta B_{ij}^n\Delta B_{kl}^n
\right\rangle
=
2D_1
\left[
(d+1)\delta_{ik}\delta_{jl}
-\delta_{ij}\delta_{kl}
-\delta_{il}\delta_{jk}
\right]\Delta t.
\label{eq:flow-increment-cov}
\end{equation}
From the predicted configuration $\rr_i^*$ we evaluate the
predicted spring forces $\FF_i^*$ and the corresponding
friction-weighted center $\Rgamma^*$. The corrected update is then
\begin{equation}
\begin{aligned}
\rr_i^{n+1}
=
\rr_i^n
&+\frac{\Delta t}{2\gamma_i}
\left(\FF_i^n+\FF_i^*\right)\\
&+\frac{1}{2}\Delta\mathbf B^n
\left[
\left(\rr_i^n-\Rgamma^n\right)
+
\left(\rr_i^*-\Rgamma^*\right)
\right]
+\Delta\rr_{i,\rm th}.
\end{aligned}
\label{eq:heun}
\end{equation}
Thus both the deterministic spring drift and the configuration-dependent
flow term are evaluated using the current and predicted configurations.
The same matrix increment $\Delta\mathbf B^n$ is used in the predictor
and corrector because both stages represent the same realization of the
random flow over a single timestep. The thermal noise is additive, so
its amplitude does not depend on the polymer configuration and requires
no predictor-corrector averaging.

\begin{table}[h]
\centering
\caption{Main parameters for the Batchelor-Kraichnan simulations.}
\label{tab:params}
\begin{tabular}{ll}
\toprule
Parameter & Value \\
\midrule
Dimension & $d=2$ \\
Chain length & $N=100$ \\
Friction ratios & $\rho=1,2,4$ \\
$k_{\rm B}T$, $\gamma_0$, $\langle |\mathbf b|^2\rangle$ & $1,1,1$ \\
Spring constant & $k=2$ \\
Time step & $\Delta t=0.01$ \\
Total simulated time & $t_{\rm tot}=10^4$ \\
Independent realizations per condition & $100$ \\
Common-flow scan & $\Wref=0,0.35,0.50,0.70,0.80,0.85$ \\
Tracked generalized modes & first 20 internal modes \\
\bottomrule
\end{tabular}
\end{table}

The simulations use the parameters listed in
Table~\ref{tab:params}. Exact equilibrium Gaussian chains are sampled
directly at $t=0$, avoiding a separate no-flow equilibration stage. Since
the equilibrium conformational distribution is independent of $\rho$,
the same ensemble of initial configurations is reused across the
different parameter conditions in order to reduce unnecessary differences arising from the initial states. Each parameter
condition nevertheless uses independent thermal-noise and random-flow
number streams, so the subsequent stochastic trajectories are not
forced by identical noise realizations. 

When the random flow is switched on, the equilibrium initial configurations are generally not representative of the driven state. We therefore use the fixed analysis cutoff
\begin{equation}
t_{\rm burn}=3\tau_1(\rho).
\label{eq:burn}
\end{equation}
This cutoff removes the initial short transient but is not assumed to guarantee convergence to the stationary driven distribution. As the slowest mode approaches the coil-stretch threshold, the relaxation time of the full extension distribution can become much longer than $\tau_1$. 

Post-burn trajectories are divided into two equal temporal halves and compared. Because Hookean polymers in random flows can develop very broad extension statistics, medians and interquartile ranges are emphasized. Means are retained as descriptive quantities but become sensitive to rare events at strong flow. Also, time samples entering the displayed extension distributions are correlated; the resulting curves are used to show distributional shape, not as independent-sample uncertainty estimates. Where uncertainty bars are shown for summary observables, they are constructed from independent realization-level statistics rather than from individual correlated time frames.

The end-to-end magnitude is normalized by the exact equilibrium median of the two-dimensional Gaussian chain. Since the end-to-end vector is Gaussian and its magnitude is Rayleigh distributed,
\begin{equation}
R_{\rm ee,eq}^{\rm med}
=\sqrt{\left\langle R_{\rm ee}^2\right\rangle\ln2}.
\label{eq:median-scale}
\end{equation}

An additional numerical check is performed. A passive infinitesimal separation vector is evolved with the same Kraichnan increments and repeatedly renormalized, providing an independent estimate of $\lambda_{\rm flow}$; the measured values track the imposed value from Eq.~\eqref{eq:flow-lyapunov}.

\section{Results in random flow}

\subsection{Global extension response and preferential excitation of slow modes}

\begin{figure}[h]
\centering
\includegraphics[width=\textwidth]{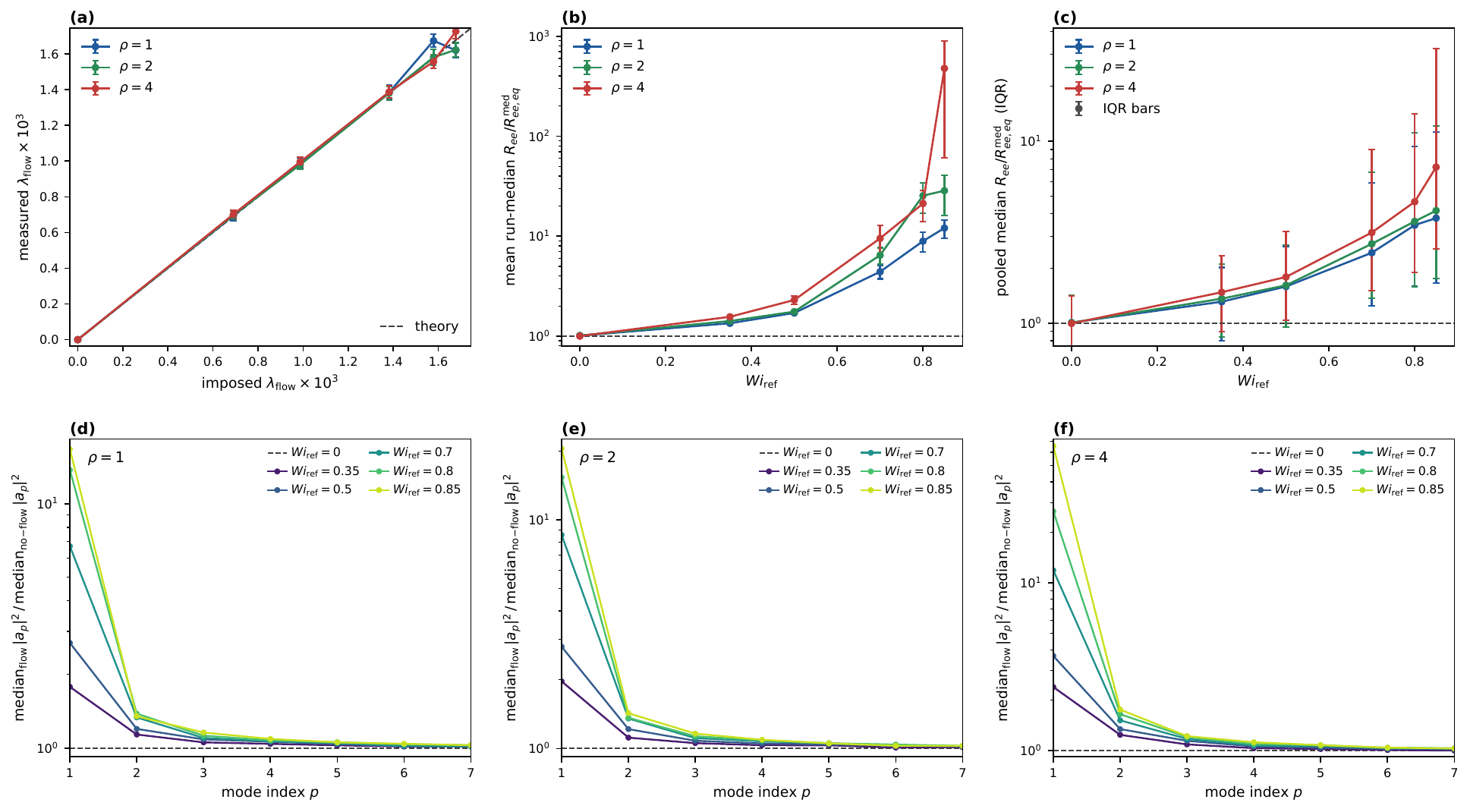}
\caption{Global extension response and selective low-mode amplification in Batchelor-Kraichnan flow for the heterogeneous Gaussian chain ($N=100$).
(a) Measured passive-separation Lyapunov exponent versus the imposed $\lambda_{\rm flow}$, validating the flow normalization.
(b) Mean of the independent-run post-burn median extension ratios $R_{\rm ee}/R_{\rm ee,eq}^{\rm med}$ versus $\Wref$; error bars denote the SEM across independent runs.
(c) Pooled post-burn median extension ratio, with bars spanning the interquartile range (IQR; 25th--75th percentiles), showing the increasing width of the extension distribution.
(d)--(f) Post-burn median generalized-mode power, normalized by the corresponding no-flow median, for the first seven modes at $\rho=1$, $2$, and $4$, respectively. Increasing $\Wref$ preferentially amplifies the slowest modes, while the higher displayed modes approach their quiescent power.}
\label{fig:global-flow}
\end{figure}

Before examining the polymer response, we independently validate the imposed Batchelor-Kraichnan flow strength using the passive-separation Lyapunov exponent. As shown in Fig.~\ref{fig:global-flow}(a), the measured growth rates closely follow the prescribed $\lambda_{\rm flow}$ over the full range considered. We then characterize the global extension through the normalized quantity
\begin{equation}
\widetilde R_{\rm ee}(t)
=
\frac{R_{\rm ee}(t)}{R_{\rm ee,eq}^{\rm med}},   
\end{equation}

where $R_{\rm ee,eq}^{\rm med}$ is the exact equilibrium median end-to-end distance. Figure~\ref{fig:global-flow}(b) shows, for each condition, the mean of the post-burn temporal medians of $\widetilde R_{\rm ee}$ computed separately for each independent run; the error bars are the SEM across those independent run-wise medians and therefore quantify run-to-run uncertainty. In Fig.~\ref{fig:global-flow}(c), by contrast, all post-burn extension samples are pooled across runs before calculating the median and interquartile range (IQR), so here the bars characterize the width of the sampled extension distribution rather than statistical uncertainty. Together, the two diagnostics show increasing extension with flow strength, accompanied by increasing run-to-run variability and distributional broadening. The response becomes particularly pronounced near the strongest-flow conditions, where the actual slowest-mode Weissenberg number
\[
\Wrho
=
\lambda_{\rm flow}\tau_1(\rho)
=
\Wref\frac{\tau_1(\rho)}{\tau_1(\rho=1)}
\]
approaches unity; for example, at $\Wref=0.85$ one obtains $\Wrho(\rho=4)\simeq0.99$. This corresponds to the regime in which flow stretching and the intrinsic relaxation of the slowest mode occur on comparable timescales. As examined explicitly in Sec.~\ref{sec:slow-convergence} below, appreciable temporal drift is already visible at $\Wref=0.70$ and becomes much stronger at $\Wref=0.80$ and $0.85$. The corresponding extension summaries should therefore be read as finite-time descriptors of the simulated window rather than as estimates of fully converged stationary extension statistics.

This broadening is accompanied by a highly selective response in generalized-mode space. For each mode $p$, we define its instantaneous modal power as the squared magnitude of the two-dimensional mode amplitude, $P_p(t)=|\mathbf a_p(t)|^2$, so that larger $P_p$ corresponds to stronger excitation of that deformation mode. Figures~\ref{fig:global-flow}(d)--\ref{fig:global-flow}(f) show the normalized post-burn median modal power
\begin{equation}
\frac{\operatorname{median}_{\rm flow}|\mathbf a_p|^2}
{\operatorname{median}_{\rm no\text{-}flow}|\mathbf a_p|^2}    
\end{equation}

for $\rho=1,2,4$. The first mode is amplified strongly as $\Wref$ increases, the second and third modes show progressively weaker enhancement, and the higher displayed modes rapidly approach their quiescent values. The same hierarchy is observed for all three friction contrasts. This behavior is naturally explained by Eq.~\eqref{eq:mode-sde}: the same multiplicative flow acts on every mode, whereas the restoring rate $\lambda_p$ increases with mode index. The flow therefore competes most effectively with relaxation in the slowest modes and probes the heterogeneous chain primarily through the low-frequency part of its relaxation spectrum. Friction heterogeneity changes that spectrum and the associated mode shapes, but the dominant random-flow response remains concentrated in the slowest modes.

\subsection{Friction heterogeneity produces a blockwise stretching asymmetry}

\begin{figure}[h]
\centering
\includegraphics[width=\textwidth]{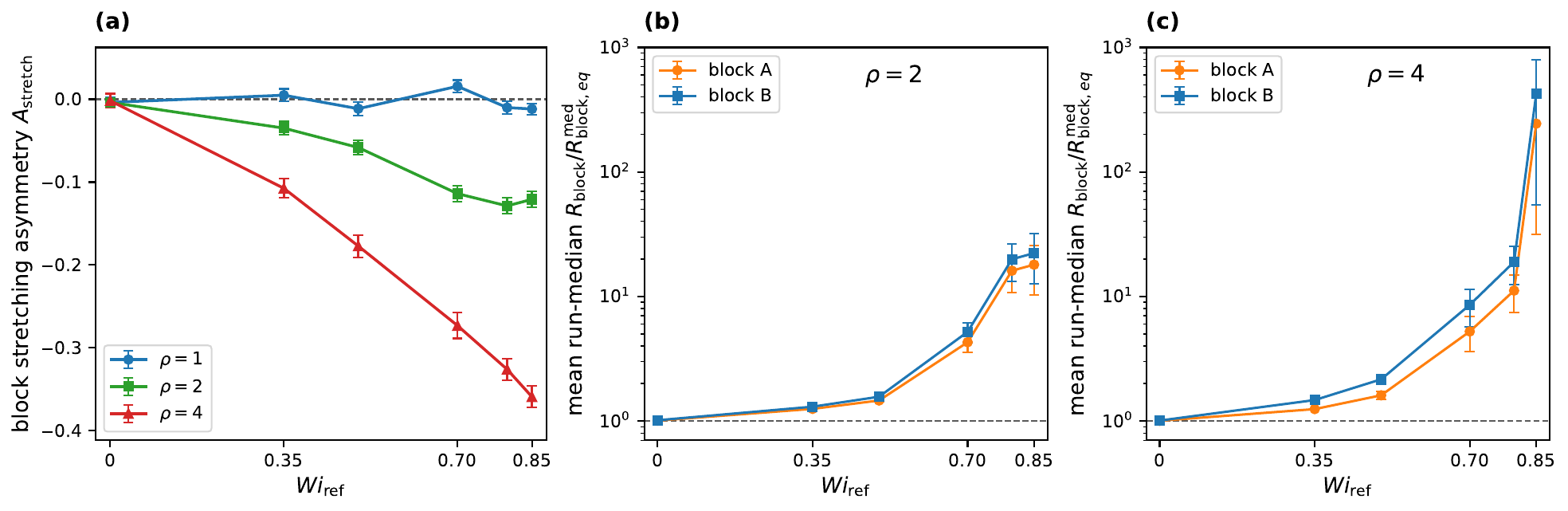}
\caption{Block-resolved stretching under a common Batchelor-Kraichnan flow.
(a) Mean post-burn block stretching asymmetry
$A_{\rm stretch}=(R_A^2-R_B^2)/(R_A^2+R_B^2)$
versus $\Wref$ for $\rho=1,2,4$; the dashed line marks equal extension of the two blocks.
Negative values indicate preferential stretching of the high-friction block B.
(b) and (c) Mean of the independent-run post-burn median block extensions,
$R_{\rm block}/R_{\rm block,eq}^{\rm med}$,
for $\rho=2$ and $\rho=4$, respectively, shown on logarithmic vertical axes; the dashed horizontal line marks the equilibrium reference value of unity.
Orange and blue denote blocks A and B. Error bars in all panels are SEMs across independent realizations.}
\label{fig:asymmetry}
\end{figure}

The central effect of the friction contrast is revealed by resolving the extension of the two halves of the chain. We define
\begin{equation}
A_{\rm stretch}
=
\frac{R_A^2-R_B^2}{R_A^2+R_B^2},
\label{eq:Astretch}
\end{equation}
where $R_A$ and $R_B$ are the end-to-end magnitudes of the two half-chains, excluding the interface bond in the corresponding block vectors. With this convention, $A_{\rm stretch}>0$ means that block A is more extended and $A_{\rm stretch}<0$ means that block B is more extended. Figure~\ref{fig:asymmetry}(a) shows the mean post-burn value of this asymmetry, with SEMs evaluated across independent realizations.

For the homogeneous chain ($\rho=1$), the measured asymmetry remains small, $|A_{\rm stretch}|\lesssim0.02$, and shows no systematic sign or monotonic trend over the flow-strength range, as expected from the exchange symmetry of the two equal halves. In contrast, $A_{\rm stretch}$ is systematically negative for $\rho=2$ and more strongly negative for $\rho=4$. At $\Wref=0.70$, for example, the mean post-burn asymmetry is approximately $-0.11$ for $\rho=2$ and $-0.27$ for $\rho=4$, while the homogeneous value remains small. The negative bias develops across the weak- to moderate-flow regime and remains clearly visible at the largest simulated $\Wref$, although its magnitude need not increase monotonically between the two strongest-flow points.

To display the underlying block extensions directly,
Figs.~\ref{fig:asymmetry}(b) and \ref{fig:asymmetry}(c)
show the two blocks separately for the moderate contrast $\rho=2$
and the stronger contrast $\rho=4$. For each block (A or B), we use
the normalized blockwise end-to-end extension
\begin{equation}
\mathcal{E}_{\rm block}
=
\frac{R_{\rm block}}
{R_{\rm block,eq}^{\rm med}},
\qquad
R_{\rm block,eq}^{\rm med}
=
\sqrt{\left\langle R_{\rm block}^2\right\rangle_{\rm eq}\ln 2},
\label{eq:block-extension}
\end{equation}
where $R_{\rm block,eq}^{\rm med}$ is the exact equilibrium median
end-to-end magnitude of either equal half-chain. The plotted values are
the means of the independent-run post-burn medians of
$\mathcal{E}_{\rm block}$, with SEMs across runs. Because the block extensions span several orders of magnitude over the
flow-strength range, panels (b) and (c) are shown on logarithmic vertical
axes. For both $\rho=2$ and $\rho=4$, the two blocks exhibit a closely tracking
overall increase with $\Wref$, reflecting the dominant common stretching
of the connected chain. At nonzero flow, however, the high-friction
B-block curve lies systematically above the low-friction A-block curve
in both panels, consistent with the negative values of
$A_{\rm stretch}$ in panel (a). The relative separation is more pronounced
for the larger friction contrast $\rho=4$, while the run-to-run
uncertainty also grows strongly at the largest flow strengths. Panels (b) and (c) therefore show that the high-friction block B is systematically more extended than the low-friction block A, even though both blocks follow the same overall increase in extension with flow strength, whereas the bounded quantity
$A_{\rm stretch}$ in panel (a) provides the clearest measure of that
relative redistribution.

The sign of the asymmetry is notable because it is opposite to the quiescent mobility asymmetry. Without flow, block A has the larger monomer MSD because it has lower friction. Under random stretching, however, the deformation bias should not be interpreted in terms of two independent block relaxation times: relaxation occurs through the global generalized modes of the connected chain. The crucial observation is that the imposed random strain preferentially amplifies the slowest mode, $p=1$, whose bead-wise shape becomes asymmetric when $\rho>1$.

This connection can be made quantitative. If a single generalized mode $p$ dominates the conformation,
\[
\rr_i-\Rgamma \simeq v_{ip}\,\mathbf a_p,
\]
then the block end-to-end vectors are proportional to
\[
\Delta v_{A,p}
=
v_{N/2,p}-v_{1p},
\qquad
\Delta v_{B,p}
=
v_{N,p}-v_{N/2+1,p},
\]
and the corresponding pure-mode stretching asymmetry is
\begin{equation}
A_{\rm stretch}^{(p)}
=
\frac{(\Delta v_{A,p})^2-(\Delta v_{B,p})^2}
     {(\Delta v_{A,p})^2+(\Delta v_{B,p})^2}.
\label{eq:pure-mode-asymmetry}
\end{equation}
For the slowest mode this gives
\[
A_{\rm stretch}^{(1)}
=
0,\;-0.216,\;-0.502
\]
for $\rho=1,2,4$, respectively. The negative simulation asymmetry therefore has the sign and characteristic scale expected directly from the heterogeneous shape of the mode that is most strongly amplified by the flow. Higher modes need not favor the same block, so the deformation bias is more precisely understood as a consequence of selective slowest-mode amplification and its asymmetric mode shape.

\subsection{Matched actual-Weissenberg control}

At fixed $\Wref$, all friction ratios are exposed to the same external random-flow intensity, since $\Wref=\lambda_{\rm flow}\tau_1(\rho=1)$ is defined using the homogeneous-chain relaxation time. The heterogeneous chains, however, have slightly different longest relaxation times $\tau_1(\rho)$, and therefore experience different chain-specific longest-mode Weissenberg numbers,
\begin{equation*}
\Wrho=\lambda_{\rm flow}\tau_1(\rho).
\end{equation*}
To determine whether the blockwise stretching asymmetry is merely a consequence of this global relaxation-time shift, we perform a separate control in which the flow strength $\lambda_{\rm flow}$ is adjusted for each friction ratio so that
\begin{equation}
\Wrho=0.70
\end{equation}
for all three chains. Using Eq.~\eqref{eq:tau-values}, the corresponding reference flow strengths are approximately $\Wref=0.7000$, $0.6780$, and $0.6006$ for $\rho=1,2,4$, respectively.

\begin{figure}[h]
\centering
\includegraphics[width=0.95\textwidth]{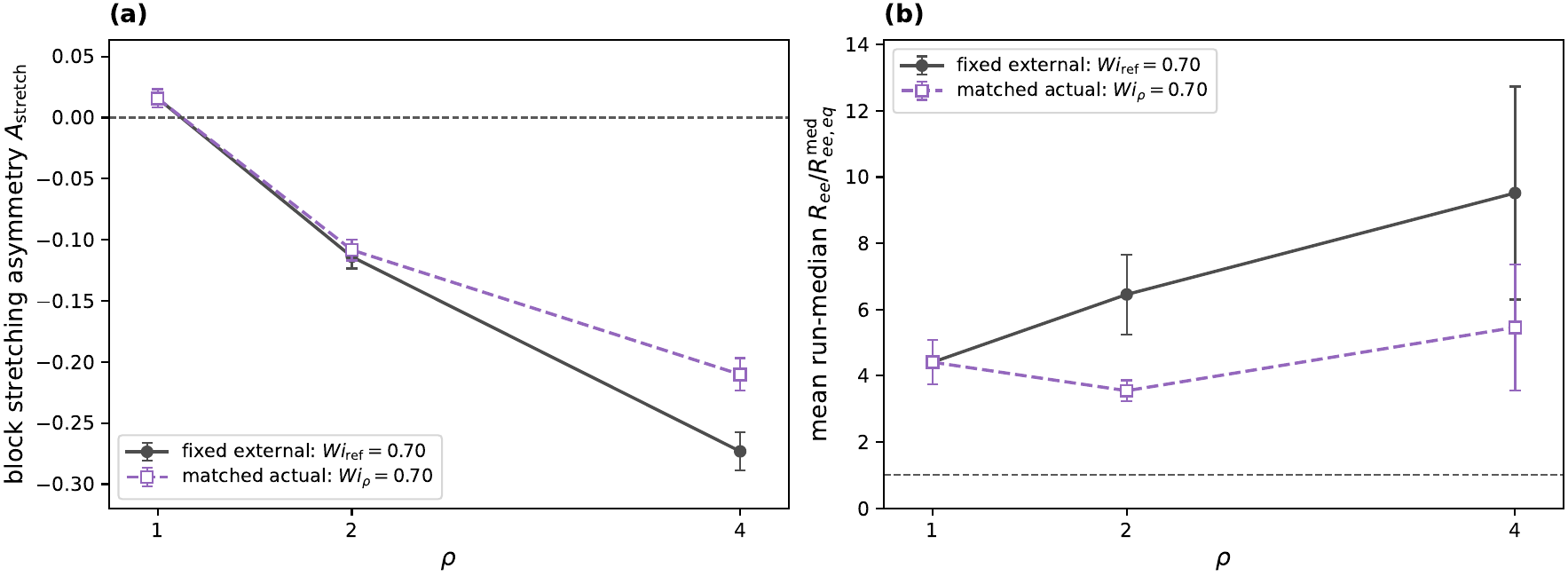}
\caption{Control at matched actual longest-mode Weissenberg number. (a) Mean post-burn block stretching asymmetry $A_{\rm stretch}$ as a function of the friction ratio $\rho$, comparing a common external flow strength, $\Wref=0.70$, with the control in which the actual longest-mode Weissenberg number is fixed at $\Wrho=0.70$ for each $\rho$. The negative asymmetry persists after the $\rho$-dependence of $\tau_1$ is compensated, although its magnitude is reduced for the largest friction contrast. The dashed horizontal line marks $A_{\rm stretch}=0$.
(b) Corresponding global end-to-end extension, shown as the mean across independent runs of the post-burn median $R_{\rm ee}/R_{\rm ee,eq}^{\rm med}$; the dashed line marks the equilibrium reference value of unity. Error bars in both panels are SEMs across independent realizations. Matching $\Wrho$ removes the simple increase of total extension with $\rho$, while the systematic blockwise deformation bias remains within the same finite-time protocol.}
\label{fig:matched-wi}
\end{figure}

Figure~\ref{fig:matched-wi}(a) compares the block stretching asymmetry obtained at the common external flow $\Wref=0.70$ with that obtained after matching the chain-specific value $\Wrho=0.70$. The asymmetry survives this matching: the homogeneous-chain value remains small, while the measured values are approximately $-0.11$ for $\rho=2$ and $-0.21$ for $\rho=4$. Matching $\Wrho$ therefore reduces part of the asymmetry at the largest friction contrast but does not remove the systematic preference for extension of the high-friction block B.

This comparison should be interpreted as a finite-time matched-$\Wrho$ control rather than as a comparison of fully converged stationary extension statistics. For the two-dimensional Kraichnan Hookean problem, the slowest-mode relaxation time near the transition is
\begin{equation}
\frac{T_{\rm rel}}{\tau_1}
\simeq
\frac{2\Wrho}{(1-\Wrho)^2},
\label{eq:kraichnan-relaxation}
\end{equation}
so at $\Wrho=0.70$ one obtains $T_{\rm rel}\simeq15.6\,\tau_1$ \cite{Celani2005}. The available post-burn windows span approximately $14$--$17\,\tau_1$ across the three friction ratios. Thus the matched control places all chains at the same distance from the slowest-mode threshold and on the same intrinsic relaxation scale, but it does not imply complete stationary convergence. What it establishes is that, under this matched finite-time protocol, the negative blockwise deformation bias persists after the simple $\rho$-dependence of the longest relaxation time has been compensated.

The corresponding global extension is shown in Fig.~\ref{fig:matched-wi}(b). At fixed $\Wref=0.70$, the total extension increases strongly with $\rho$, whereas after matching $\Wrho$ this simple monotonic trend disappears and the run-to-run uncertainty becomes substantial, particularly for $\rho=4$. Together, these results show that friction heterogeneity produces a robust redistribution of deformation between the two blocks, whereas the total chain extension is more strongly influenced by the $\rho$-dependence of the longest relaxation time.

\subsection{Strong stretching and slow approach to stationarity}
\label{sec:slow-convergence}

Because the Hookean chain is infinitely extensible, random stretching produces increasingly broad extension statistics as the slowest mode approaches the coil-stretch threshold. Importantly, all conditions simulated here remain on the coiled side of that threshold: the largest chain-specific value is $\Wrho\simeq0.99$ for $\rho=4$ at $\Wref=0.85$. The Hookean Kraichnan model therefore still possesses a normalizable stationary distribution at every simulated parameter value, although Eq.~\eqref{eq:mode-tail} shows that its large-extension tail becomes extremely broad and that low-order moments can already diverge well below $\Wrho=1$.

\begin{figure}[h]
\centering
\includegraphics[width=0.9\linewidth]{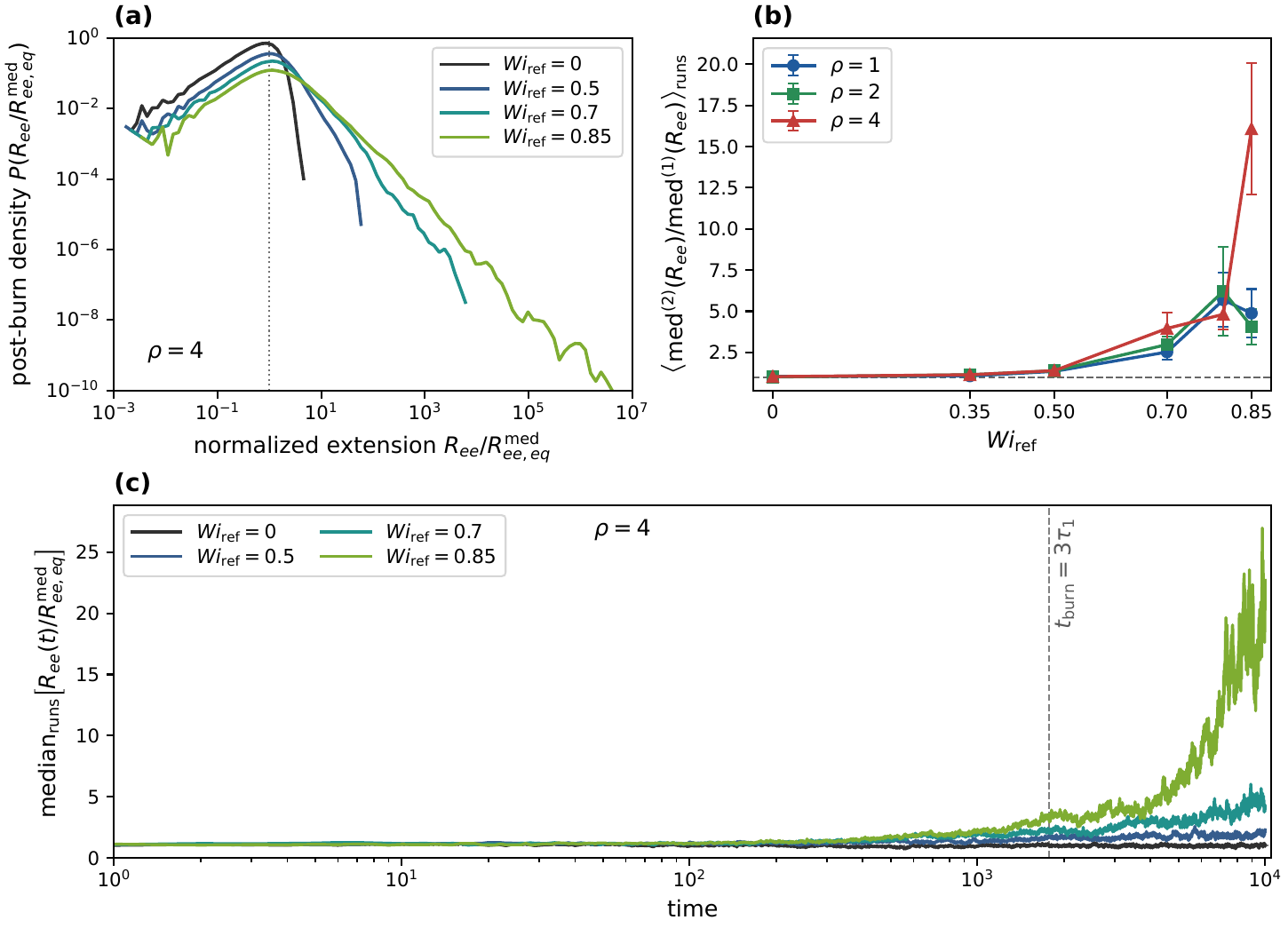}
\caption{Strong stretching and slow convergence toward the stationary Hookean regime.
(a) Post-burn extension densities for $\rho=4$ at selected flow strengths, with extension normalized by the exact equilibrium median $R_{\mathrm{ee,eq}}^{\mathrm{med}}$; the dotted vertical line marks the equilibrium-median scale.
(b) Mean ratio $\langle \mathrm{med}^{(2)}(\Ree)/\mathrm{med}^{(1)}(\Ree)\rangle_{\rm runs}$, where the superscripts denote the first and second halves of the post-burn interval; error bars are SEM across independent runs and the dashed line marks unity.
(c) Time-resolved median normalized extension for $\rho=4$ at the same selected flow strengths, with the dashed vertical line indicating the fixed analysis cutoff $t_{\rm burn}=3\tau_1$.
Increasing flow produces progressively broader extension statistics and increasingly slow convergence over the available simulation window. The observed drift is therefore interpreted as a finite-time relaxation effect below the Hookean coil-stretch threshold, rather than as an intrinsic loss of stationarity.}
\label{fig:stationarity}
\end{figure}

Figure~\ref{fig:stationarity}(a) shows the post-burn extension distributions for $\rho=4$ at selected flow strengths. In the absence of flow the distribution is concentrated around the equilibrium scale, whereas increasing $\Wref$ produces progressively broader distributions with increasingly long tails toward large extension. The broadening becomes especially pronounced as $\Wrho$ approaches unity.

To diagnose whether the simulated time window has reached the stationary regime, we divide the post-burn interval of each realization into two equal halves and compare their median end-to-end extensions. Figure~\ref{fig:stationarity}(b) shows
\begin{equation}
\left\langle
\frac{\mathrm{med}^{(2)}(\Ree)}
     {\mathrm{med}^{(1)}(\Ree)}
\right\rangle_{\rm runs},
\end{equation}
where the superscripts $(1)$ and $(2)$ denote the first and second halves of the post-burn interval. A value close to unity indicates little systematic finite-time drift. The ratio is close to unity at zero and weak flow but becomes appreciably larger than unity already around $\Wref=0.70$, with still stronger departures at $\Wref=0.80$ and $0.85$.

The slow convergence is expected from the known relaxation spectrum of the Kraichnan Hookean problem. In two dimensions, the slowest-mode relaxation estimate in the near-threshold regime is
\begin{equation*}
\frac{T_{\rm rel}}{\tau_1}
\simeq
\frac{2\Wrho}{(1-\Wrho)^2}.
\end{equation*}

For example, for $\rho=4$ at $\Wref=0.70$ one has $\Wrho\simeq0.816$, giving $T_{\rm rel}\simeq48\,\tau_1$, whereas the available post-burn interval is only about $13.9\,\tau_1$. The observed temporal drift is therefore consistent with critical slowing of convergence toward the stationary heavy-tailed distribution.

The corresponding time dependence for $\rho=4$ is shown directly in Fig.~\ref{fig:stationarity}(c). At weak flow the typical extension is comparatively stable after $t_{\rm burn}$, whereas at larger $\Wref$ it continues to evolve throughout the post-burn interval. Accordingly, extension statistics for $\Wref\geq0.70$ are interpreted here as finite-time summaries rather than fully converged stationary estimates. This caveat does not originate only at the largest flow strength and does not imply that the subcritical Hookean model lacks a stationary distribution. The central blockwise deformation bias is already visible at weaker flow strengths. Broad stationary power-law tails and the strong divergence of relaxation times near the coil-stretch transition are established properties of Hookean polymers in smooth random flows \cite{Balkovsky2000,Celani2005,Celani2006}.

\section{Discussion}

The quiescent and flow-driven results show two distinct consequences of the same friction contrast. In the absence of flow, lower friction produces faster local motion, so block A has the larger early- and intermediate-time MSD. The equilibrium structure, however, is unchanged, and translational diffusion is determined by the total friction. The heterogeneous chain is therefore not a copolymer with different elastic properties, but an otherwise uniform Gaussian chain with spatially varying dissipation.

Under random straining, local mobility and flow-induced deformation need not favor the same block. The random velocity gradient preferentially amplifies the slowest generalized mode, while friction heterogeneity changes the bead-wise shape of that mode. For $\rho>1$, the resulting $p=1$ mode shape carries a negative pure-mode block asymmetry, quantitatively accounting for the direction and characteristic magnitude of the observed deformation bias. The matched-$\Wrho$ control shows that this bias is not accounted for solely by the $\rho$-dependence of the longest global relaxation time. The mechanism is therefore a global mode-shape effect of the connected heterogeneous chain.

Although the simulations use an equal contiguous diblock, several analytical results are more general. For any fixed multiblock friction profile, the friction-weighted translational diffusion coefficient retains the scaling $D_\gamma\propto N^{-1}$, with a prefactor determined by the block fractions and frictions. More generally, the generalized eigenproblem and the projected Batchelor-Kraichnan mode dynamics remain valid for any positive diagonal friction profile. Piecewise-constant multiblock chains also inherit a hierarchy of local spectral cutoffs, $\lambda_{\max,\alpha}=4k/\gamma_\alpha$. What is not established analytically here is the detailed partitioning of extension among more than two blocks; that depends on the resulting global mode shapes and remains to be tested beyond the diblock simulations.

The mechanism may be relevant whenever different portions of a polymer experience different local dissipation while being subjected to a common fluctuating deformation field. Polymer blends can generate heterogeneous mobilities
\cite{Niedzwiedz2007}, while crowded or disordered media can produce heterogeneous polymer reaction kinetics \cite{KwonSung2019}. Random and turbulent flows likewise produce intermittent stretching histories governed by competition between deformation and polymer relaxation \cite{Chertkov2000,Celani2005,Gerashchenko2005,Picardo2023}. The present model isolates one specific ingredient within that broader setting: a spatially organized friction contrast along a single chain.

Several limitations are deliberate. The polymer is two-dimensional and Gaussian, with no excluded volume, and the Hookean springs are infinitely extensible. Below the coil-stretch threshold the Hookean model possesses a stationary distribution but develops increasingly broad algebraic tails, while above the threshold no normalizable stationary Hookean extension distribution exists. A finitely extensible nonlinear elastic (FENE) spring law or another finite-extension force is therefore required to obtain physically bounded stretching and a stationary description beyond the Hookean coil-stretch threshold. Hydrodynamic interactions between beads are also neglected, although they can modify turbulent stretching statistics \cite{Ganesh2026}. The Batchelor-Kraichnan field is also smooth across the entire chain and delta-correlated in time, so it represents synthetic fluctuating small-scale strain rather than a direct numerical simulation of Navier-Stokes turbulence.

Natural extensions include finite-extension springs, excluded volume, hydrodynamic interactions, three-dimensional chains, finite-correlation-time random flows, and eventually resolved turbulent velocity fields. On the heterogeneity side, unequal block fractions, multiple blocks, continuous or quenched random friction profiles, and correlations between friction and elasticity would test how broadly the deformation-partitioning mechanism persists. The present Gaussian diblock provides a controlled baseline in which the effect of friction heterogeneity can be separated from these additional ingredients.

\section{Conclusions}

We have studied a two-dimensional Gaussian Rouse chain with blockwise friction heterogeneity and uniform elasticity, first in quiescent conditions and then in an incompressible Batchelor-Kraichnan random flow. The underlying continuous-time no-flow Langevin model preserves the equilibrium Gaussian conformation, reproduces the expected local mobility asymmetry, yields the exact $D_\gamma\propto N^{-1}$ translational scaling for fixed multiblock friction profiles, and develops a friction-dependent generalized relaxation spectrum. 

In random flow, the principal result is a systematic spatial asymmetry of deformation: the high-friction block becomes more stretched than the low-friction block. The asymmetry strengthens with friction contrast, develops through the weak- to moderate-flow regime, and remains visible at the strongest finite-time conditions. It also persists under a matched chain-specific longest-mode Weissenberg-number control, showing that the observed bias is not accounted for solely by the change in the longest global relaxation time. Generalized-mode analysis provides a quantitative mechanism: the random strain preferentially amplifies the slowest mode, and the heterogeneous bead-wise shape of that mode directly predicts the sign and characteristic scale of the blockwise asymmetry.

All simulated chain-specific longest-mode Weissenberg numbers remain below the Hookean coil-stretch threshold. As this threshold is approached, the stationary extension distribution develops very broad algebraic tails and the relaxation time grows rapidly, producing substantial temporal drift over the finite simulation window. Extension statistics at the larger flow strengths are therefore interpreted as finite-time summaries of slow convergence rather than as evidence for a loss of stationarity. Overall, the results show that spatial friction heterogeneity alone can redistribute deformation along an otherwise mechanically uniform polymer.

\section*{Data and Code Availability}

The simulation code, data, and generated figures associated with this work are available on GitHub at \url{https://github.com/arpand2004/blockwise-friction-kraichnan-polymer} and archived on Zenodo at \url{https://doi.org/10.5281/zenodo.22857749}.

\section*{Conflict of Interest}

The author declares no competing financial or non-financial interests.

\end{document}